\documentclass[journal]{IEEEtran}

\pdfoutput=1
\usepackage{soul}
\usepackage[pdftex]{graphicx}
\usepackage{color}
\usepackage{multicol} 
\usepackage{array} 
\usepackage{wasysym}
\usepackage{hyperref}
\usepackage{eurosym}
\usepackage{pifont}
\usepackage{subfigure}
\usepackage[bottom]{footmisc}
\usepackage{lineno}
\usepackage{listings}

\usepackage{lineno}

\begin{document}
\bstctlcite{IEEEexample:BSTcontrol}

\title{ProtoDUNE-DP Light Acquisition and Calibration Software}

\author{D.~Belver, J.~Boix, E.~Calvo, C.~Cuesta, A.~Gallego-Ros, I.~Gil-Botella, S.~Jim\'{e}nez, C.~Lastoria, T.~Lux, I.~Mart\'{i}n, J.J.~Mart\'{i}nez, C.~Palomares, J.~Soto-Oton, A.~Verdugo

\thanks{D.~Belver, E.~Calvo, C.~Cuesta, A.~Gallego-Ros, I.~Gil-Botella, S.~Jim\'{e}nez, C.~Lastoria, I.~Mart\'{i}n, J.J.~Mart\'{i}nez, C.~Palomares, J.~Soto-Oton, and A.~Verdugo are with Centro de Investigaciones Energ\'{e}ticas, Medioambientales y Tecnol\'{o}gias (CIEMAT), 28040 Madrid, Spain. J.~Boix and T.~Lux are with Institut de F\'{i}sica d\'{}Altes Energies (IFAE) - The Barcelona Institute of Science and Technology, Campus UAB, 08193 Bellaterra (Barcelona), Spain.}

\thanks{Manuscript received March 2021.\textit{ (Corresponding authors: C.\,Cuesta and S.~Jim\'{e}nez)} }}



\maketitle

\begin{abstract} 

ProtoDUNE-DP is a 6$\times$6$\times$6\,m$^3$ liquid argon time-projection-chamber operated at the CERN Neutrino Platform in 2019-2020 as a prototype of the Dual Phase concept for the DUNE Far Detector. The Photon Detection System (PDS) is based on 36 8-inch photo-multiplier tubes (PMTs) and allows triggering on the scintillation light signals produced by cosmic rays and other charged particles traversing the detector. The acquisition and calibration software specifically developed for the ProtoDUNE-DP PDS is described in this paper. This software controls the high-voltage power supplies, the calibration system, and the PDS DAQ. It has been developed with Qt Creator, and features different operation modes, and a graphical user interface. This software has already been validated and used during the ProtoDUNE-DP operation.
\end{abstract}


\IEEEpeerreviewmaketitle

\section{Introduction} 
\label{sec1}


The Deep Underground Neutrino Experiment (DUNE) aims at address key questions in neutrino physics and astroparticle physics~\cite{DUNE_LBL, DUNE_SN, DUNE_BSM}. DUNE will consist of a near detector placed at Fermilab close to the production point of the muon neutrino beam of the Long-Baseline Neutrino Facility (LBNF), and four 10\,kt fiducial mass liquid argon time-projection chambers (LArTPCs) as far detector in the Sanford Underground Research Facility (SURF) at 4300\,m.w.e. depth at 1300\,km from Fermilab~\cite{DUNEtdrv2,DUNEtdrv4}. 

In order to gain experience in building and operating such large-scale LAr detectors, two prototypes are operated at the CERN Neutrino Platform with the specific aim of validating the design, assembly, and installation procedures, the detector operations, as well as data acquisition, storage, processing, and analysis. The two prototypes employ LArTPCs as detection technology. One prototype only uses LAr, called ProtoDUNE Single-Phase (SP)~\cite{ProtoDUNESP}, and the other uses argon in both its gaseous and liquid state, thus the name ProtoDUNE Dual-Phase (DP)~\cite{wa105,Cuesta:2019yeh}. Both detectors have similar sizes of $\sim$700\,t.

ProtoDUNE-DP has an active volume of 6$\times$6$\times$6 m$^{3}$ corresponding to an active mass of 300\,t.  In ProtoDUNE-DP charged particles crossing the detector are ionizing the argon. The electrons released in the ionization drift towards the anode due to an electric field, the drift field, applied between anode and cathode. The ionization charge is then extracted, amplified, and detected in gaseous argon above the liquid surface allowing a low energy threshold with high signal to-noise ratio, and thus a good pattern reconstruction of the events. Ionization processes in LAr produce ionization electrons and scintillation photons. The scintillation light signal is used as trigger for non-beam events, to determine precisely the event time, and there is also a possibility to perform calorimetric measurements and particle identification. Two Cosmic Ray Tagger (CRT) planes were added to the external wall of the ProtoDUNE-DP cryostat to trigger on muon-tracks passing through both CRTs.

ProtoDUNE-DP DAQ is composed by two independent DAQ systems: one for the charge signals collected in the anode and the second one for the light signals from the Photon Detection System (PDS). Data taken with both DAQ systems are synchronized by means of a White Rabbit (WR) network\footnote{https://white-rabbit.web.cern.ch/}.

LAr scintillation light is in the far vacuum ultraviolet, having a wavelength centered at 127\,nm~\cite{Heindl:2010zz}. The prompt scintillation light (usually referred to as S1 signal) in LAr has two components: S1 fast with a lifetime of $\sim$6\,ns, and S1 slow of $\sim$1.6\,$\mu$s. In addition, the electro-luminiscence secondary scintillation light, called S2, is produced in the gas phase of the detector when electrons, extracted form the liquid, are accelerated in the electric field between the liquid phase and the anode. To obtain the information about the the particle which caused the ionization and the LAr purity contained in the S1 signal, the pulse shape information is needed and this determines the sampling rate. The S2 signal contains information about the track topology and the digitization window length is given by the extreme case of the drift time of electrons from the cathode to the anode. The expected time scale of the S2 in ProtoDUNE-DP is of the order of hundreds of microseconds. Then, in order to not loose information, a sampling rate  at least tens of ns is required and a digitization window of at least 4 ms are required to the light acquisition system.

The PDS of ProtoDUNE-DP~\cite{protoDUNElight} is formed by 36 8-inch cryogenic photo-multiplier tubes (PMTs), R5912-02MOD from Hamamatsu~\cite{protoDUNEPMTs, Belver:2020qmf}, placed below the cathode grid. As the PMTs are not sensitive to 127\,nm light, a wavelength shifter is placed on top of the PMTs to convert this light to the visible range where PMTs are sensitive. The PDS facilitates the detection of the primary scintillation signals, S1, which provide the absolute time reference for the interaction events. It also enables recording of the S2 electro-luminescence signals. 

A light calibration system (LCS) has been developed for ProtoDUNE-DP to monitor the PMT performance and obtain an equalized PMT response~\cite{Belver:2019lqm}. In the light calibration system, the light source is provided inside a black box by blue LEDs of 465\,nm using Kapustinsky circuits~\cite{1985NIMPA.241..612K} as LED driver, and transmitted by a fiber system ending with a fiber pointing at each PMT. The LEDs are pulsed at 1\,kHz with 30\,ns pulse width. Inside the black box there are six LEDs placed in a hexagonal geometry and a SiPM (Silicon Photo-Multiplier) at the center to check the LEDs performance. The direct light from each LED goes to a fiber, and the stray light to the SiPM used as reference sensor.

The acquisition software is designed to cope with the different data taking modes. This implies the implementation of 3 trigger modes: 
\begin{itemize}
\item Light trigger: This trigger is based on the coincidence of selected PMT signals (over a threshold level) during a time window. The trigger rate is on the Hz-kHz range depending on the threshold and the number of PMTs requested to pass the threshold in coincidence.
\item External trigger: In this case the PDS receives an external trigger signal from the DAQ global computer at 10\,Hz or from the CRT planes at around 0.3\,Hz.
\item Calibration mode: A external trigger signal (NIM) is received from the light calibration system at 1\,kHz synchronized with the calibration light pulse sent to the PMTs. In this mode it is also possible to take data with random trigger at a configurable rate if all light sources are turned off.
\end{itemize}

\begin{figure*}[ht]
    \centering
    \includegraphics[width=0.95\textwidth]{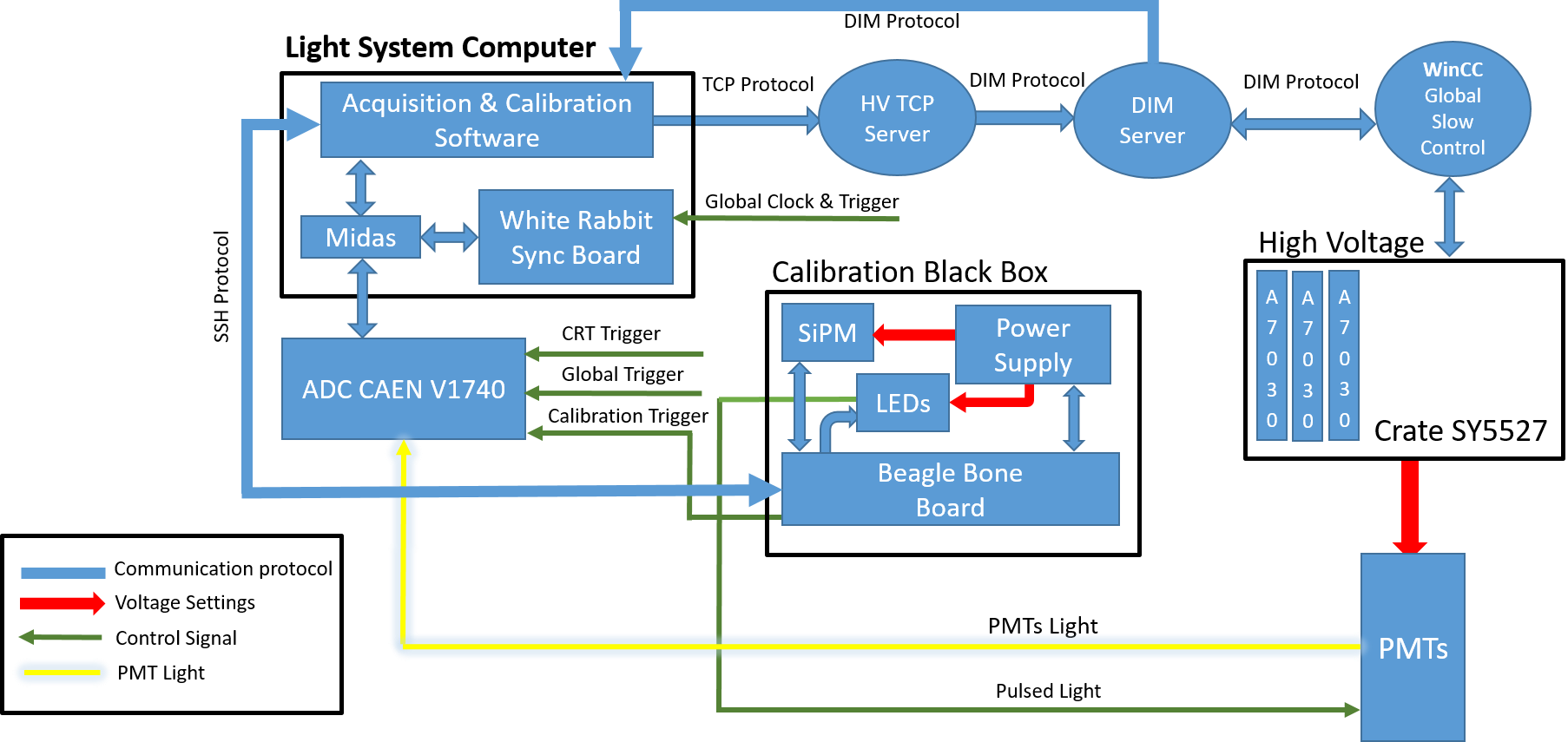}
    \caption{ProtoDUNE-DP light readout block diagram. All systems and protocols presented are involved in the calibration of the PMTs and in the acquisition of the PMTs signal during the detector operation.}
    \label{fig:diagram}
\end{figure*}

Such an experimental system requires a dedicated software that meets all the different needs
. This paper describes the light data acquisition and calibration software (ACS) specifically developed for ProtoDUNE-DP. The light ACS allows the user to choose the acquisition trigger mode, control and define the acquisition settings (front-end and high voltage), select the light trigger, arrange the calibration mode and settings, and provide the graphical user interface. The diagram of the ProtoDUNE-DP light readout is displayed in Fig.~\ref{fig:diagram} showing all systems and protocols involved in the PMT calibration and acquisition of the PMT signals during the detector operation. The systems controlled by the ACS are the PMT high-voltage, the calibration black box, and the light DAQ, and are described in section~\ref{sec2}. The ACS is detailed in section~\ref{sec3}, where the operation modes, the graphical user interface and the execution flow are described. The light data obtained with the ACS in ProtoDUNE-DP are presented in section~\ref{sec4}.

The environment chosen to develop the light ACS is Qt Creator\footnote{https://www.qt.io/product/development-tools}, and the programming language used is C++. Qt Creator allows to create quick and easy graphical user interfaces with a drag and drop environment, it has multiple and powerful software development kits and any C++ library can be easily integrated. Qt is a cross-platform environment, but the drivers required to run the SPEC card (for the connection to the WR network) made Linux Ubuntu 14.04 LTS the chosen operating system.


\section{Controlled systems} 
\label{sec2}
In this section the different systems related to the ProtoDUNE-DP light ACS will be defined at a hardware level. The different protocols used to communicate with them and how these protocols are coded to establish the necessary control actions involved in the PMT operation will be described. The basic PMT operation actions are, biasing the PMTs, controlling the LCS and managing the data taking. A class is defined for each system. These classes contain data structures to store and manage the parameters for their configuration and methods to manage the different communication protocols.

\subsection{High-voltage} 
PMTs are biased with three CAEN A7030\footnote{http://www.caen.it/} modules with twelve channels each. They are inserted in a SY5527 power supply system from CAEN that is controlled using WinCC\footnote{https://new.siemens.com}, a control and data acquisition interface system from Siemens. The DIM\footnote{https://dim.web.cern.ch} (Distributed Information Management) protocol is used as the communication layer with the power supply as with the rest of slow control systems in ProtoDUNE-DP, and also in ProtoDUNE-SP~\cite{SinglePhaseControl}. All the operations executed by WinCC are transparent to the light acquisition software.
 
\subsubsection{DIM Protocol}  
\begin{figure}[ht]
    \centering
    \includegraphics[width=0.45\textwidth]{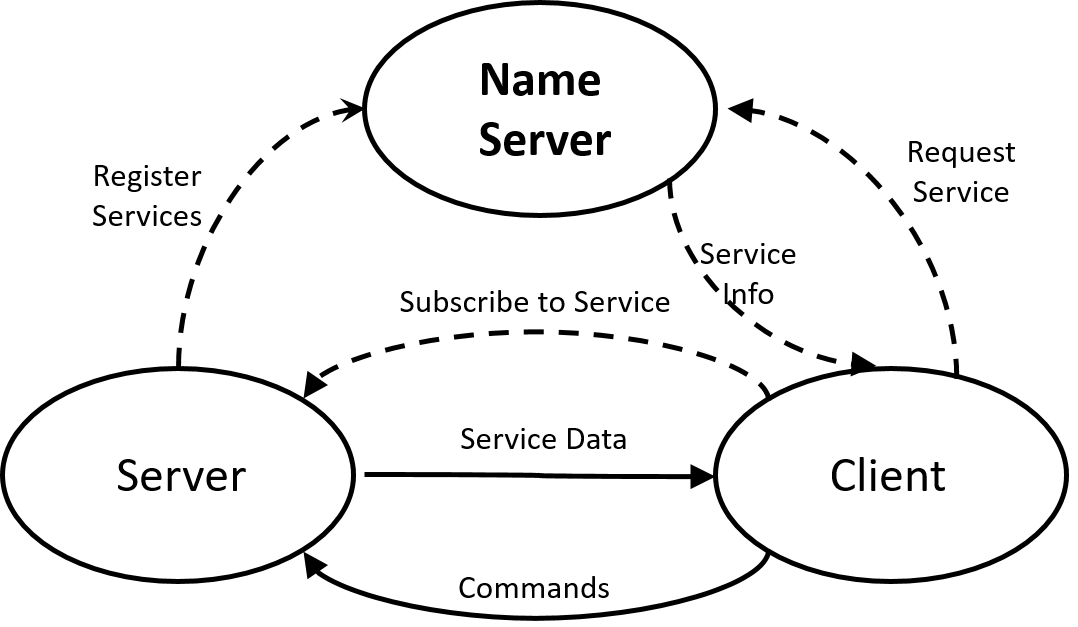}
    \caption{DIM Operation Diagram and data flow. Figure taken from~\cite{GASPAR2001102}.}
    \label{fig:DIM}
\end{figure}

It was designed at CERN as a multipurpose communication layer with the aim of unifying the requirements and constraints for the different detector operations. DIM is based on the client/server paradigm in which a service is a set of data recognized by a name. Servers publish their services by registering them with the name server and clients subscribe to services by asking the corresponding server name and then interacting directly with that server. Figure~\ref{fig:DIM} displays the control and data flow among the basic components of a DIM system. The name server receives service registration messages from servers and service requests from clients. Once a client obtains the service information from the name server it can then subscribe to services or send commands directly to the server~\cite{GASPAR2001102}.

\subsubsection{Data structures} 
The status of each voltage channel is defined in a data structure with two elements: one float member to store the voltage value, and a binary to store the channel on/off status. The values to be programmed on the power supply are stored using this structure type, with each channel stacked in a dynamic vector. When the  control system reads the status of the power supply, it is stored in the same kind of dynamic vector with a structure, so the desired and current values can be compared during ramp-up and ramp-down operations.

\subsubsection{Classes and Methods}  
Two classes are used to communicate with the high voltage system. In the first one, services with the desired voltage values are registered. It also allows to receive the channel values in real time subscribing to the name server published services that have been read by WinCC directly from the high voltage board. The methods defined in this class wrap the server and client libraries provided by DIM to adapt them to our data structures. The second class is defined for establishing a bridge between the machine that is running the light acquisition program and a CentOS server capable to register services to the name server. This bridge must be implemented to avoid an incompatibility between the DIM protocol and the services published from an Ubuntu system.

When the high-voltage values need to be updated, the data are sent to a TCP server that manages messages with an event driven technique based on signals and slots\footnote{https://doc.qt.io/qt-5/signalsandslots.html}. If new data arrive, a signal is triggered executing a slot that publishes the values using the methods defined in the class wrapping DIM.

The main program is capable of subscribing to the high voltage values from the name server directly with DIM, without intermediary protocol, so the power supply status can be known on demand.

Since in the detector more sensors are installed which emit level of light which might damage the PMTs (the LEDs of the cameras intalled to inspect the interior of the detector and the lamps of the purity monitors), an interlock mechanism has been established over DIM protocol. It avoids turning the PMT high-voltage on when the systems are operating and vice-versa.

\subsection{Calibration black box} 
The calibration black box is the core of the LCS providing the control of the light source for the PMT calibration~\cite{Belver:2019lqm}. The black box operation and communication are controlled by a BeagleBone Black\footnote{https://beagleboard.org/} board, an open-hardware mini computer that runs an embedded Debian Linux distribution.

The Black box architecture has two basic groups of elements: optical and control elements. The optical ones, inside the box, are 6 LEDs and the SiPM reference sensor. These are assembled in a 3D printed structure with a central part containing the SiPM and control PCB, and 6 arms to which each a LED driver PCB is fixed. For each LED a printed board circuit has been designed. It implements a Kapustinski~\cite{1985NIMPA.241..612K} configuration to get few-nanosecond pulses. The control PCB has two main purposes. First, it distributes the LED bias voltages to the different driver PCBs as also the signals which trigger the light pulse emitted by the LEDs. Second, it duplicates the SiPM signals for testing purposes, allowing to digitize them with the CAEN ADC and, at the same time, to integrate the SiPM signal and digitize the output with the built-in ADC of the BeagleBone Black module. The programmable real time units (PRU) of the BeagleBone Black are used to provide the trigger signals and to reset the integrator for data taking.

Low level software is in charge of controlling the PRUs. The first one controls all the hardware and passes the data through shared memory to the second PRU, achieving a very precise timing. The second one is in charge of communication with the main processor. For the main processor a set of instructions with the basic functionalities has been developed. These have been attached by Secure Shell (SSH) commands. Python scripts on the BeagleBone are used to control the voltage settings of the SiPM and the LED bias voltages. 

\subsubsection{Communication protocol} 
SSH is a very natural way to communicate with a BeagleBone board. To implement the SSH communication, the library libssh\footnote{https://www.libssh.org/} has been wrapped using a class to manage communications. Libssh is a multiplatform C library that has been imported into the Qt solution. This library implements the SSHv2 protocol on client and server side, so we can remotely control the BeagleBone board with a set of custom shell script commands that have been developed for this purpose.

\subsubsection{Data structures}
All features needed to control the operation of the light source are grouped in a data structure. The collection of commands provides the ability to set the values stored in this data structure into the calibration black box system. Those values are the reference sensor and LEDs bias voltage, the selection of active LEDs, the LED pulses frequency, the number of pulses, and the possibility to inhibit or activate the trigger sent to the digitizer board.

\subsubsection{Classes and methods}
Control of a Linux-based system is straightforward using commands sent over SSH. The class programmed for this purpose provides methods to establish the communication and to send the desired commands to perform the configuration, to start a process on the BeagleBone and receive its response to manage the execution of the control operation in progress. 


\subsection{Light DAQ} 
The light readout front-end is based on the commercial ADC V1740\footnote{https://www.caen.it/products/v1740/} from CAEN. This 12-bit VME digitizer has 64 analog input channels with 2~Vpp dynamic range and a maximum sampling rate of 62.5~MS/s. 
 In the ADC module, settings as for example trigger thresholds are not set per channel but the 64 channels are organized internally in groups of 8 channels. As soon as the input of one of the 8 channels fulfills the threshold condition, the whole group is flagged as having been triggered. By setting a specific coincidence level, it is possible to require that at least that specified number of groups have been flagged as triggered to start the event acquisition. Besides the trigger on the signal amplitude, also an external trigger signal, either NIM or TTL logic, can be connected to the ADC
This module is connected to the light readout computer by an 80~MB/s optical link (CAEN CONET2).

The light readout DAQ system is based on MIDAS, a DAQ framework developed by PSI and TRIUMF\footnote{https://midas.triumf.ca}, which was also successfully used for the light readout of the 4-ton DP LArTPC demonstrator~\cite{311}. MIDAS performs the communication with the ADC through the CAEN CONET2 link for configuration and data acquisition and also takes care of the storage of the ADC received data on disk files.

The light ACS acts like a top abstraction layer to the MIDAS DAQ framework, which provides the user with an interface designed to configure and control the data acquisition.

As the light readout data is acquired and stored separately from the anode charge data, to enable the possibility to synchronize both data acquisitions, a White Rabbit (WR) SPEC card\footnote{https://white-rabbit.web.cern.ch/} is installed on the light DAQ computer. This card generates a time-stamp on reception of the global trigger signal. This time-stamp is incorporated in the light readout data stream for each trigger to enable the matching with the charge data.

\subsubsection{Interface with MIDAS} 
MIDAS is running in the same computer as the ACS, so it is possible to interact with it directly with system commands and disk files. Qt native class QProcess\footnote{https://https://doc.qt.io/qt-5/qprocess.html} is used to send the system command that loads the configuration into MIDAS and enables the system to start and stop the acquisition.

\subsubsection{Data structures} 
Two data structures are defined to contain all the necessary parameters to configure the digitizer during data acquisition. The first structure stores the number of events to be taken. A binary dynamic vector stores the groups of channels to set active during the acquisition. This data structure also includes the sampling rate, the number of samples that defines the window length, and finally the buffer size that defines the board data throughput and allows to optimize the acquisition speed depending on the window size.

With the second data structure all trigger related parameters are configured. The first parameter is the trigger source, that, as mentioned before, it can come from the internal PDS DAQ trigger, from the global trigger computer, from the CRT panels DAQ or from the calibration system. An internal trigger can be configured with a given threshold over the PMT acquired signal. If the internal trigger is selected, the channels that are going to participate in building the trigger are stored in a dynamic vector. A different threshold for each group of channels can be set with the board. This set of trigger levels are stored in its corresponding dynamic vector. The post-trigger  parameter allows to define the fraction of acquisition window to be displayed after the trigger. Another parameter is needed to set the time to wait for another trigger. The last parameter in the trigger configuration indicates the desired number of trigger groups flagged during the trigger window to launch an acquisition event. 

\subsubsection{Classes and methods} 
The class designed to manage MIDAS provides the necessary methods to configure and control MIDAS software that will define the digitizer behaviour.

To configure MIDAS a script file is built with the necessary parameter values contained in the data structures described in the previous section. This script is loaded into MIDAS database before the data acquisition is performed, as well as the slow control parameters, high voltage and calibration black box configuration. The database is stored with the acquired data so the full system configuration for each run can be traced back in the data analysis.

After storing the desired configuration parameters the class is equipped with methods to start and stop the acquisition when requested during the program control flow. The system response is captured in a string to parse the response enabling the program to monitor and control the data acquisition performance and give feedback to the program user.

\section{Light acquisition and calibration software} 
\label{sec3}

After describing the involved systems and the strategies implemented to control them, the interface and the features offered to the user are presented in this section.

\subsection{Operation Modes} 
In order to help the user and simplify the PDS management, three different operation modes are implemented. With this approach only the features needed for each specific operation mode are available. 
When the ACS starts the main window offers the program user the three operation options.

The first mode, called \textit{Calibration Mode}, gives the user the capability to set the active high voltage channels and the desired voltage value for each channel. It also gives full control of the calibration black box. In addition, it is possible to set the LEDs OFF so the user can run the acquisition controlling the trigger frequency and acquire data in random trigger mode. On the data acquisition side, the calibration mode grants access to some of the front-end settings. The program user can choose the channel groups to be acquired, the sampling period, the window size and the post-trigger samples. This mode is used to determine the PMT gain at the operating HV.

The second operation mode, called \textit{Gain vs Voltage}, is designed to perform a high voltage scan and acquire the PMTs signals at each selected voltage. In this way, the user can set a start and end voltage values with a step voltage in between them, choose any configuration for the calibration black box and the same features available for the digitizer displayed in \textit{Calibration Mode}. This mode is required to obtain the calibration constants and determine the PMT gain at any given HV. 

The third operation mode is the \textit{Acquisition mode}, developed to acquire regular data. It provides access to all the features the ADC V1740 board offers. The PMTs can be biased individually and the window size and sampling rate set to the desired values. 
\begin{figure}[ht]
    \centering
    \includegraphics[width=0.35\textwidth]{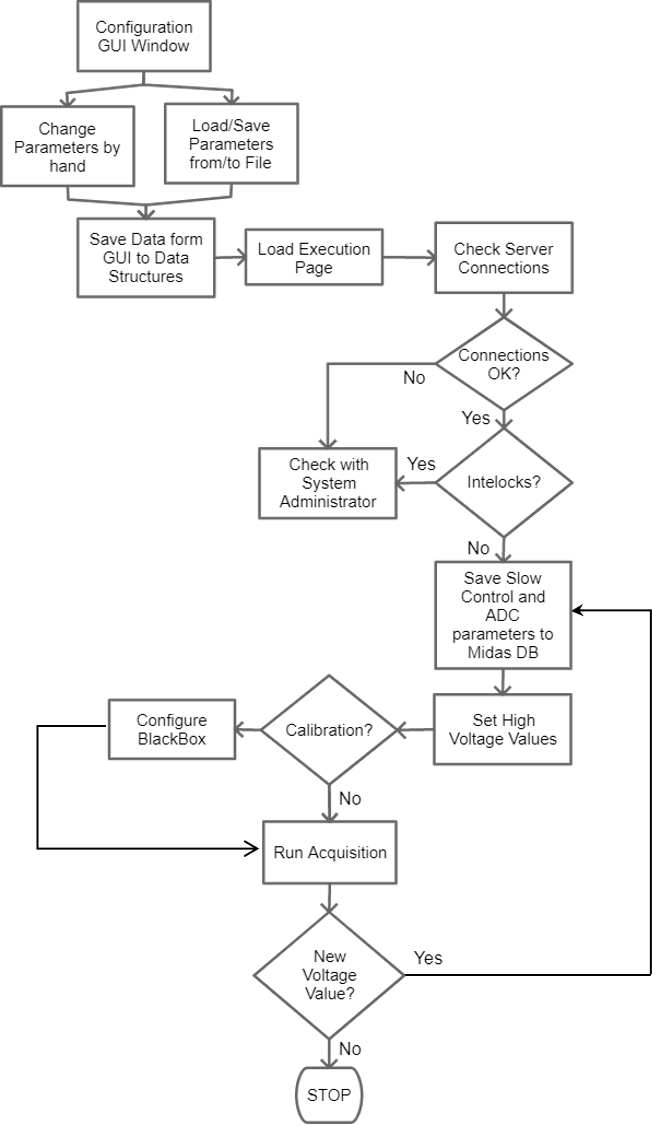}
    \caption{Execution flow diagram for the three operation modes.}
    \label{fig:flow}
\end{figure}
\subsection{Graphical User Interface} 

\begin{figure*}[ht]
    \centering
    \includegraphics[width=0.75\textwidth]{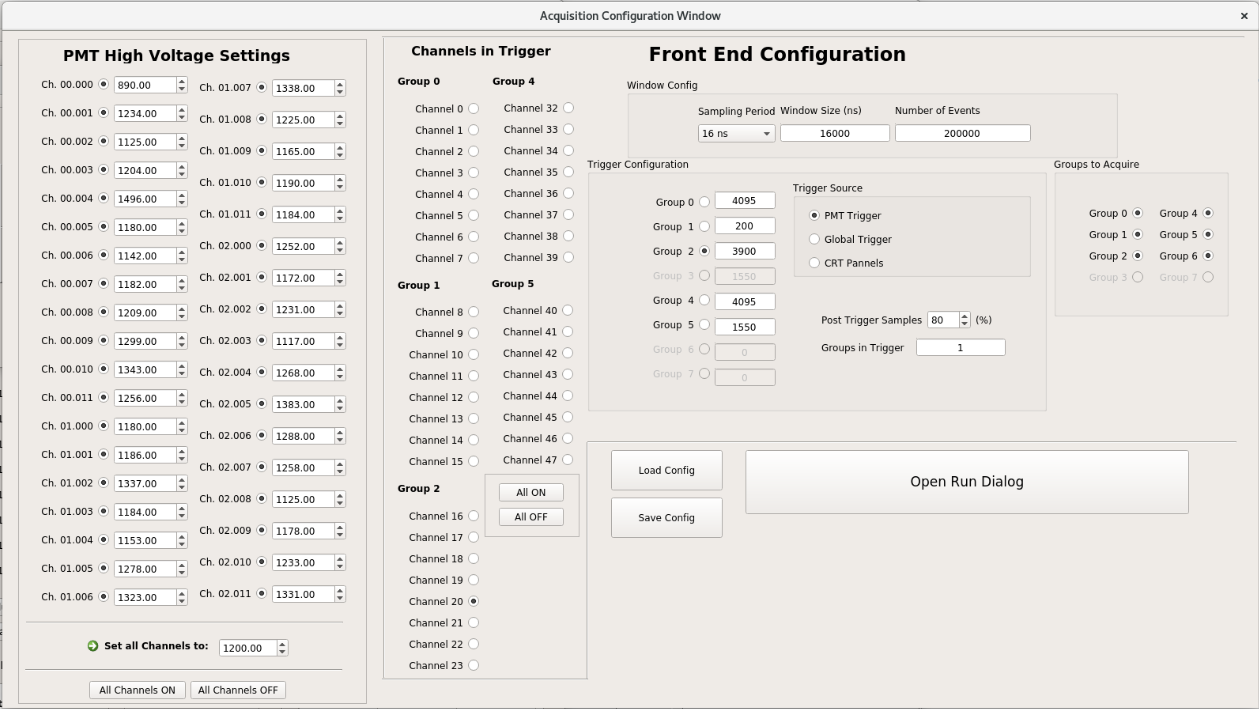}
    \caption{Acquisition configuration window of the ACS. On the left panel the voltage and status can be chosen for each channel. The next panel on the right enables the user to select which PMT will take part on the PMT Trigger. On the upper right corner, the sampling rate, the length of the acquisition window, and the number of events can be selected. Below this block there is the trigger source selection and, if PMT trigger is selected, the block beside it permits to select which group will contribute to the trigger generation and its correspondent threshold values in ADC counts (0-4095). On the right the ACS user can choose which PMT group is going to be acquired.}
    \label{fig:Config}
\end{figure*}

For each of the three operation modes, two windows are designed offering the parameters that can be tuned for each mode. The first window is the configuration window where the user can set the parameters for the chosen operation mode or load them from a configuration file. An example of the acquisition configuration window is shown in Figure~\ref{fig:Config}.

\begin{figure*}[ht]
    \centering
    \includegraphics[width=0.75\textwidth]{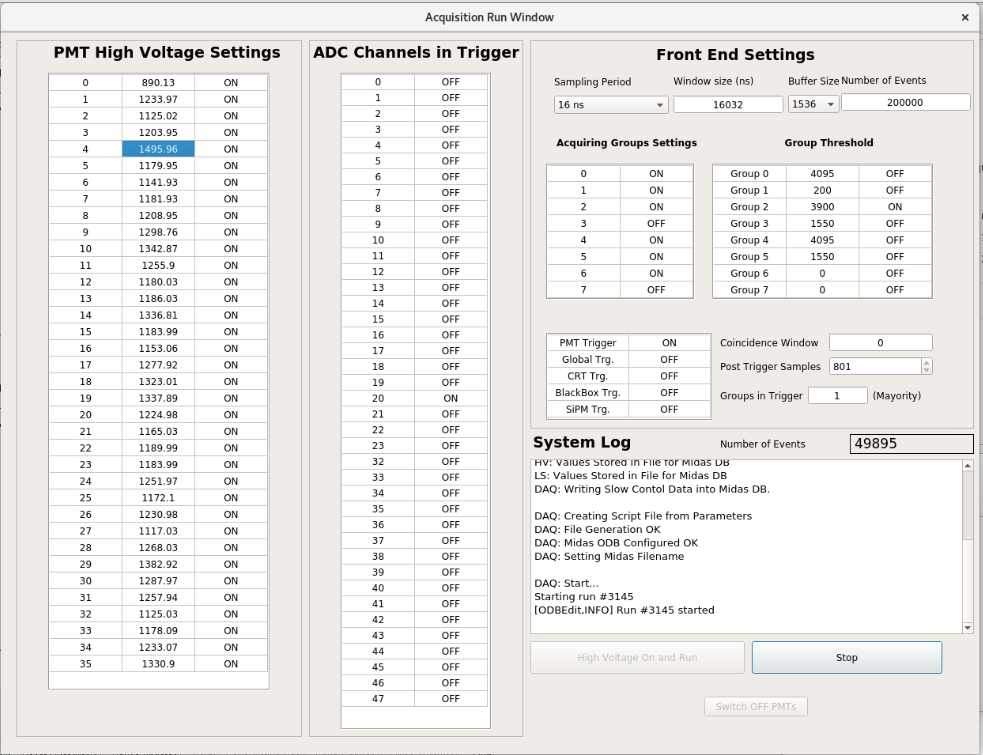}
    \caption{Acquisition execution window. On the left block the PMT voltages are displayed. The second column shows the values the user has set on the configuration window, and once the acquisition is in process it will present the values in real time during the HV ramp up or ramp down. The rest of the panels are similar to the ones described in the configuration window but in this occasion they are only informative, no changes can be done. The user is informed in real time of the number of events acquired. The system Log window shows the status of the processes in execution tagged accordingly to the system they interact with.}
    \label{fig:Run}
\end{figure*}

Once the parameters are set, the execution window is prompted. This second window displays informatively all the configuration parameters chosen on the previous window, so the user can check once more everything is correct before pressing the Run button, that initialises all the actions developed to operate the PMTs. This window shows, in real time, all the operations performed during the execution described in the next section. While the PMT data is being acquired, it can be visualized using the event display described before. The user can select which and how many channels are displayed.  An example of the acquisition execution window is shown in Figure~\ref{fig:Run}.

\begin{figure}[ht]
    \centering
    \includegraphics[width=0.45\textwidth]{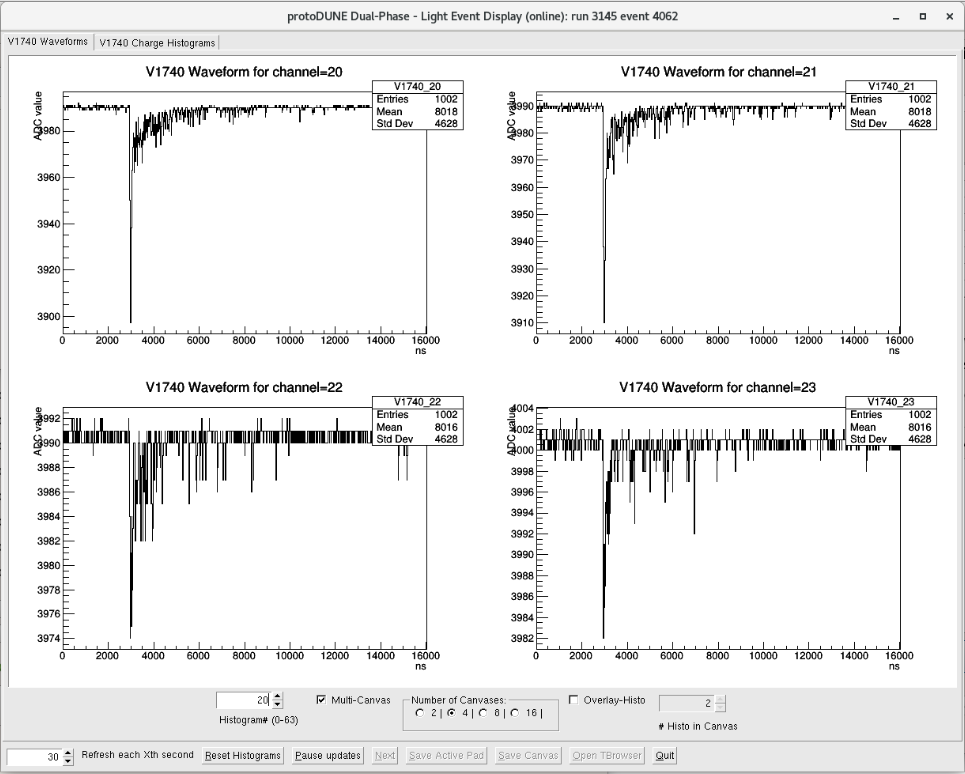}
    \caption{Event display with four PMT channels selected showing a ProtoDUNE-DP light event with a S1 signal. The ADC value is shown in the y-axis and the time in the x-axis with 16 ns sampling time. }
    \label{fig:AnaDisplay}
\end{figure}

To monitor the data acquisition, an event display was developed based on the ROOTANA package\footnote{https://midas.triumf.ca/MidasWiki/index.php/ROOTANA}. A ProtoDUNE-DP light event example is shown in Figure~\ref{fig:AnaDisplay}. It runs online as the data is taken, using the interfaces classes provided by MIDAS and adapted to our ADC model. It can also run offline taking previously stored MIDAS files as an input. The event display has the capability to show the raw waveform signals acquired from the PMTs and the integrated signal in a second tab, calculated online from the raw waveform.

\subsection{Execution Flow} 
Modularity and code reuse are the base of the development of the ACS. The three operation modes do not only share the same classes but the execution flow is also similar to all of them. 

The user selects an operation mode and the processes described in the diagram shown in Figure~\ref{fig:flow} are executed, starting with the configuration window where the user can modify or load from a file  the configuration parameters. Next, the user can save the configuration or open the Run dialog. The program stores then the data from the configuration window to the data structures and checks the connectivity with the DIM server, with the TCP server and with the calibration black box. If all communications are correct, it loads the settings on the execution window for a final check. All operations are displayed on the System Log window specifying to which interface the operation being executed belongs to.  

If no correction needed, the user can click the \textit{Start Acquisition} option and the program checks if there is a conflict with the cameras or the purity monitors. If not, the configuration is saved to MIDAS database and the DAQ is configured. All this information is stored alongside the acquired PMTs waveforms. Then the PMTs are biased and voltage values are displayed in the High Voltage Settings section during the ramp up. When they reach their nominal values, they are compared with the user settings to find out if there is any inconsistency with the desired settings. Then the data acquisition will start unless one of the calibration modes are selected branching to the black box configuration and then the data acquisition starts. On a 
\textit{Gain vs Voltage} test, a voltage settings and acquisition loop is carried out until the maximum voltage is reached. The acquisition stops once the desired number of events is reached or if the program user aborts the execution.

\section{Data taking} 
\label{sec4}

The ProtoDUNE-DP photon detection system has been operative during more than a year (July 2019 - September 2020) taking data with the ACS previously described. Data have been taken almost every day during short time periods (typically $\sim$1-2 hours/day). Most of the time, data taking took place remotely. PMT HVs had to be switched off during the operation of camera LEDs and purity monitors to avoid any damage on the PMTs. Longer time periods of light data ($\sim$12 hours/run) were possible, typically at night, proving the stability of the software. User feedback was used for improving the software, for instance, the number of acquired events is shown in the acquisition execution window and the system log is saved in a file per request of the users. A total of 86 million events have been acquired with a total duration of 366 hours. 

During the PMT operation, the obtained data is first stored locally split in 2.1\,GB files to facilitate its handling. At the end of every day a script is scheduled to convert the data into ROOT\footnote{https://ph-root-2.cern.ch/} format. The ROOTANA package is used to convert the binary data in the MIDAS file in a ROOT file for data analysis. After that, both ROOT and MIDAS files are sent to CASTOR\footnote{http://castor.web.cern.ch/castor/} storage system at Fermilab and EOS\footnote{http://information-technology.web.cern.ch/services/eos-service} storage system at CERN. Local data is removed from the light system acquisition computer after the backup operation. This process is done offline once per day.

\section{Conclusions}

The acquisition and calibration software (ACS) specifically developed for ProtoDUNE-DP photon detection system has been described. It controls three different subsystems: the PMT high-voltage, the light calibration system (LCS) and the light DAQ. The development for the ProtoDUNE-DP light ACS is an important step for the final DUNE detector software. 

The goal of every user interface is to make things easy for the person in charge of managing the systems controlled by the program. The presented system achieves this goal, bringing together the control of very different systems into one tool that can have the photon detection system running and supervised.

It is also important to remark that the designed software hierarchy allows to easily reuse existing classes to control similar systems or easily adapt them to manage different ones. Any other system can be controlled with DIM, adapting the data structure to the new control variables and renaming the services and clients to have the new communication interface ready to run. Adapting the SSH class is also possible programming on the remote Unix based system a set of commands, or using Unix standard commands if no custom action is needed. Regarding MIDAS, the developed class programs the database and gives basic orders so, if the digitizer is changed, adapting the data structure and the script generation to the new board specifications, would give control of the new board.

The ProtoDUNE-DP light acquisition and calibration software has been validated and used for more than a year (July 2019 - September 2020) taking data without major problems.

\section*{Acknowledgment}

This project has received funding from the European Union Horizon~2020 Research and Innovation programme under Grant Agreement no.~654168; from the Spanish Ministerio de Economia y Competitividad (SEIDI-MINECO) under Grants no.~FPA2016-77347-C2-1-P, FPA2016-77347-C2-2-P, MdM-2015-0509, and SEV-2016-0588; from the Comunidad de Madrid; and the support of a fellowship from ''la Caixa" Foundation (ID 100010434) with code LCF/BQ/DI18/11660043. IFAE is partially funded by the CERCA program of the Generalitat de Catalunya

\bibliographystyle{IEEEtran}
\bibliography{biblio}

\begin{thebibliography}{10}
\providecommand{\url}[1]{#1}
\csname url@samestyle\endcsname
\providecommand{\newblock}{\relax}
\providecommand{\bibinfo}[2]{#2}
\providecommand{\BIBentrySTDinterwordspacing}{\spaceskip=0pt\relax}
\providecommand{\BIBentryALTinterwordstretchfactor}{4}
\providecommand{\BIBentryALTinterwordspacing}{\spaceskip=\fontdimen2\font plus
\BIBentryALTinterwordstretchfactor\fontdimen3\font minus
  \fontdimen4\font\relax}
\providecommand{\BIBforeignlanguage}[2]{{%
\expandafter\ifx\csname l@#1\endcsname\relax
\typeout{** WARNING: IEEEtran.bst: No hyphenation pattern has been}%
\typeout{** loaded for the language `#1'. Using the pattern for}%
\typeout{** the default language instead.}%
\else
\language=\csname l@#1\endcsname
\fi
#2}}
\providecommand{\BIBdecl}{\relax}
\BIBdecl

\bibitem{DUNE_LBL}
B.~Abi and Aothers, ``Long-baseline neutrino oscillation physics potential of
  the dune experiment,'' \emph{EPJC}, vol.~80, no.~10, p. 978, 2020.

\bibitem{DUNE_SN}
\BIBentryALTinterwordspacing
B.~Abi \emph{et~al.}, ``{Supernova Neutrino Burst Detection with the Deep
  Underground Neutrino Experiment},'' \emph{EPJC}, 2020. [Online]. Available:
  \url{arXiv:2008.06647}
\BIBentrySTDinterwordspacing

\bibitem{DUNE_BSM}
\BIBentryALTinterwordspacing
B.~Abi \emph{et~al.} (2020) {Prospects for Beyond the Standard Model Physics
  Searches at the Deep Underground Neutrino Experiment}. [Online]. Available:
  \url{arXiv:2008.12769}
\BIBentrySTDinterwordspacing

\bibitem{DUNEtdrv2}
\BIBentryALTinterwordspacing
B.~Abi \emph{et~al.} (2020) {Deep Underground Neutrino Experiment (DUNE), Far
  Detector Technical Design Report, Volume II DUNE Physics}. [Online].
  Available: \url{arXiv:2002.03005}
\BIBentrySTDinterwordspacing

\bibitem{DUNEtdrv4}
B.~Abi \emph{et~al.}, ``{Volume IV. The DUNE far detector single-phase
  technology},'' \emph{JINST}, vol.~15, no.~08, p. T08010, 2020.

\bibitem{ProtoDUNESP}
B.~Abi \emph{et~al.}, ``{First results on ProtoDUNE-SP liquid argon time
  projection chamber performance from a beam test at the CERN Neutrino
  Platform},'' \emph{JINST}, vol.~15, no.~12, p. P12004, 2020.

\bibitem{wa105}
\BIBentryALTinterwordspacing
L.~Agostino \emph{et~al.} (2014) {LBNO-DEMO: Large-scale neutrino detector
  demonstrators for phased performance assesment in view of a long-baseline
  oscillation experiment, CERN-SPSC-2014-013, SPSC-TDR-004}. [Online].
  Available: \url{arXiv:1409.4405}
\BIBentrySTDinterwordspacing

\bibitem{Cuesta:2019yeh}
\BIBentryALTinterwordspacing
C.~Cuesta, ``{Status of ProtoDUNE Dual Phase},'' in \emph{{2019 European
  Physical Society Conference on High Energy Physics}}, 10 2019. [Online].
  Available: \url{arXiv:1910.10115}
\BIBentrySTDinterwordspacing

\bibitem{Heindl:2010zz}
T.~Heindl \emph{et~al.}, ``{The scintillation of liquid argon},'' \emph{EPL},
  vol.~91, no.~6, p. 62002, 2010.

\bibitem{protoDUNElight}
C.~Cuesta \emph{et~al.}, ``{Photon detection system for ProtoDUNE dual
  phase},'' \emph{JINST}, vol.~12, p. C12048, 2017.

\bibitem{protoDUNEPMTs}
D.~Belver \emph{et~al.}, ``{Cryogenic R5912-20Mod Photomultiplier Tubes
  Characterization for the ProtoDUNE Dual Phase Experiment},'' \emph{JINST},
  vol.~13, p. T10006, 2018.

\bibitem{Belver:2020qmf}
D.~Belver \emph{et~al.}, ``{First testing of the Hamamatsu R5912-02Mod
  photomultiplier tube at 4-bar pressure and cryogenic temperature},''
  \emph{JINST}, vol.~15, no.~09, p. P09023, 2020.

\bibitem{Belver:2019lqm}
D.~Belver \emph{et~al.}, ``{A Light Calibration System for the ProtoDUNE-DP
  Detector},'' \emph{JINST}, vol.~14, no.~04, p. T04001, 2019.

\bibitem{1985NIMPA.241..612K}
J.~S. {Kapustinsky} \emph{et~al.}, ``{A fast timing light pulser for
  scintillation detectors},'' \emph{NIMA}, vol. 241, pp. 612--613, Dec. 1985.

\bibitem{SinglePhaseControl}
A.~Kehrli \emph{et~al.}, ``{The ProtoDUNE Single Phase detector control
  system},'' \emph{EPJ Web Conf.}, vol. 214, p. 01024, 2019.

\bibitem{GASPAR2001102}
C.~Gaspar \emph{et~al.}, ``{DIM, a portable, light weight package for
  information publishing, data transfer and inter-process communication},''
  \emph{Computer Physics Communications}, vol. 140, no.~1, pp. 102 -- 109,
  2001, {CHEP2000}.

\bibitem{311}
B.~Aimard \emph{et~al.}, ``{A 4 tonne demonstrator for large-scale dual-phase
  liquid argon time projection chambers},'' \emph{JINST}, vol.~13, p. P11003,
  2018.

\end{thebibliography}

\end{document}